\documentclass[aps,pra,reprint,groupedaddress,amsmath,amssymb,preprintnumbers,showpacs,longbibliography]{revtex4-1}
\usepackage{graphicx}
\usepackage{bm}
\usepackage{dcolumn}
\newcommand{\eq}[1]{Eq.~(\ref{#1})}

\newcommand{\be}{\begin{equation}}
\newcommand{\ee}{\end{equation}}
\newcommand{\ba}{\begin{eqnarray}}
\newcommand{\ea}{\end{eqnarray}}
\newcommand{\bs}{\begin{subequations}}
\newcommand{\es}{\end{subequations}}
\newcommand{\bw}{\begin{widetext}}
\newcommand{\ew}{\end{widetext}}

\usepackage[final]{showlabels}
\usepackage{xcolor}

\begin{document}

\title{Nondipole effects in tunneling ionization by intense laser pulses}

\author{Lars Bojer Madsen}
\affiliation{Department of Physics and Astronomy, Aarhus
University, DK-8000 Aarhus C, Denmark}

\date{\today}

\begin{abstract}
The limit of decreasing laser frequency can not be considered independently from nondipole effects due to increase in the laser-induced continuum electron speed in this limit. Therefore, in this work, tunneling ionization in the adiabatic limit is considered for an effective field that includes effects beyond the electric-dipole term to first order in $1/c$, with $c$ the speed of light. The nondipole term describes the interaction resulting from the electric dipole-induced velocity of the electron and  the magnetic field component of the laser. The impact of this term on the ionization rate, tunnel exit point, momentum at the tunnel exit and electron dynamics is discussed. In the appropriate limit, the results of a nondipole strong-field approximation approach and those of the strict adiabatic limit, where time and field strength are parameters, are consistent. The nondipole strong-field approximation approach is used to identify nonadiabatic modifications of  the initial conditions. The results open up an avenue to include nondipole effects in the initial tunneling ionization step in semiclassical models of strong-field and attosecond physics. 
\end{abstract}

\maketitle

\section{\label{sec:introduction} Introduction}
Single ionization of atoms and molecules is the initial key process that triggers  a range of dynamics in strong-field and attosecond physics~\cite{Krausz2000,Krausz2009}. The force on the electron from the field may steer it back to scatter on the parent ion. This laser-driven electron motion may lead to high harmonic generation,  high-energy above-threshold ionization or multiple ionization~\cite{Corkum1993}. The initial ionization step is also key in the interpretation of, e.g.,  attoclock experiments~\cite{Eckle2008,Pfeiffer2012,Kheifets2020},  as well as in time-resolved perspectives on laser-induced electron diffraction~\cite{Zuo1996,Blaga2012} and strong-field holography~\cite{Huismans2011,CarlaFaria2020}. 

When the angular frequency, $\omega$, of the laser pulse is much smaller than the ionization potential, $I_p$, (atomic units (a.u.) are use throughout) the active electron adjusts to the instantaneous value of the electric field such that ionization may be described by static tunneling at an instant in time. This regime is refereed to as the strict adiabatic linit, and this is the regime of main consideration in this work. In this regime time and field strength of the laser pulse are parameters in the modelling of the ionization process. As discussed below, the treatment of this limit is complicated by the fact that it can not be considered independently from a consideration of nondipole effects~\cite{Reiss2008,Reiss2014}.

Setting aside for a moment these latter nondipole-related complications, a typical way to proceed is as follows. After the  adiabatic tunneling step, field-induced dynamics take place. These dynamics can be captured by classical considerations~\cite{Corkum1993}, and, when including phases along the trajectories, can be used for analysis of interference structures in photoelectron momentum distributions~\cite{Li2014,Nikolay2016}. As a result, in the adiabatic regime, semiclassical simulation models, where the first step is quantum mechanical tunneling and the second is classical propagation from the exit point (a concept justified, e.g.,  by  Bohmian analysis~\cite{Ivanov2017})  have belonged among the most used approaches in studying strong-field phenomena.  This approach is often referred to as the two-step model when used for ionization. When high harmonic generation is considered, one uses the word three-step model, where, in the third step, the electron recombines. Tunneling, being the first strep in this adiabatic regime, is central in these models and for this reason there is interest in applying analytical tunneling rate formulas for atoms~\cite{LL,Smirnov1966,PPT,ADK}, and molecules~\cite{MOADK,Murray2011,WFAT}. Modifications of tunneling by permanent dipoles~\cite{Holmegaard2010,WFAT}  and polarizabilities~\cite{Brabec2005,Pfeiffer2012,Maurer2012} have also been considered. 

Along with the rate, the initial conditions at the tunnel exit are central for the semiclassical simulation. At the tunnel exit, tunneling theory predicts a Gaussian distribution in the momentum transverse to the polarization direction~\cite{Delone1991}, when the field is not too strong~\cite{Batishchev2010}, and predicts a vanishing momentum in the longitudinal field (polarization) direction.  In particular the latter initial condition has caused some challenges when pursuing agreement between semiclassical models and experimental data. For example, when modelling strong-field ionization at near-infrared fields, it was found that a nonvanishing value for  the longitudinal momentum along the laser polarization direction could sometimes improve the agreement between the semiclassical two-step model and experiment~\cite{PfeifferPRL2012,Sun2014,Eckart2018}.  The theoretical justification for such a nonvanishing longitudinal initial momentum is nonexisting in the strict adiabatic tunneling limit, i.e., when tunneling is considered to occur in a field that can be considered as static from the point of view of the much faster electronic time-scale. However, analytical approaches, which include nonadiabatic effects in the tunneling step, such as the strong-field or Keldysh-Faisal-Reiss approximation~\cite{Keldysh1964,Reiss1980,Faisal1973},   give predictions for nonvanishing initial momenta along the polarization direction at the tunnel exit when $\omega$ is nonvanishing; see the review \cite{Popruzhenko2014} for a very thorough discussion of the Keldysh theory including discussions of the physical nature of exit points and initial momenta at these. See also Refs.~\cite{Teeny2016,Ni2016,Xu2017,Xu2018,Ni2018,Ni_PRA2018} for related theory discussions.

 In the present work, the situation is elucidated by  showing that a nonvanishing longitudinal momentum along the laser polarization direction at a nondipole-modified tunnel exit emerges as a consequence of an effective nondipole field component along the propagation direction in the strict adiabatic tunneling limit.  According to the Keldysh criterion, the tunneling picture becomes appropriate as $\gamma = \omega \kappa / F_0 \ll 1$, for $\kappa^2/2= I_p$ and $F_0$ the field strength~\cite{Keldysh1964}. In this adiabatic limit, one could expect the tunneling description to become increasingly accurate as $\omega$ decreases. However, the limit of decreasing $\omega$ cannot be considered independently of nondipole effects~\cite{Reiss2008,Reiss2014}. Namely, when $\omega$ decreases, the quiver velocity of the electron, $F_0/\omega$, increases and the nondipole effect due to the magnetic field component of the laser pulse cannot be neglected~\cite{Reiss2008,Reiss2014}.  At first glance, the magnetic component impedes a description in terms of an effective electric dipole-type interaction and this could be a reason why analytical descriptions including nondipole effects have applied the strong-field approximation (SFA) approach, see, e.g., Refs.~\cite{Klaiber2013,Lein2018PRA,Lein2019PRA,Ni2020,Lein2021,Lund2021}. 
 
 It is the purpose of this work to deal with nondipole effects in tunneling caused by intense fields ($\sim 10^{14}$ W/cm$^2$) and to discuss how tunneling concepts can be applied for a suitably defined effective nondipole field in the nonrelativistic, adiabatic regime described by
\be
F_0/c \ll \omega \ll I_p/2.
\label{condition}
\ee
The latter inequality ensures that tunneling at the instantaneous field strength is acccurate, the factor of $1/2$ is due to the nondipole correction discussed below. The former inequality ensures that nondipole effects are captured by expansion of the vector potential to first order in $1/c$ and that the nondipole SFA Hamiltonian is accurate~\cite{JensenPRA2020}.  Equation \eqref{condition} applies to atoms and molecules for midinfrared fields at intensities covered by laser sources that are currently being developed~\cite{Wolter2015}, and where nondipole effects have been observed~\cite{Keller2014,Willenberg2019}. In this regime, access to analytical approaches is important for interpretation of data showing signatures of nondipole effects. Note in passing that nondipole effects have also been observed with near-infrared fields~\cite{Smeenk2011,Haram2019,Doerner2019,Hartung2021}; see Refs.~\cite{Wang2020,Haram2020,Maurer2021} for recent reviews on nondipole effects in intense laser pulses. At near-infrared wavelengths the adiabaticity condition in \eq{condition} is fulfilled to lesser degree than at midinfrared fields and therefore the concept of tunneling at an instantaneous field strengths is less valid.

The paper is organized as follows. In Sec.~II, the theory and the results are presented and discussed.  Section III gives the conclusion and an outlook.

\section{Theory and discussion}

The present analysis will be based on an approximate description of the nondipole effects to order $1/c$. In the high-intensity, high-frequency regime~\cite{shakeshaft,kylstra,Forre2014,Forre2015,Forre2016a,Forre2016b,Moe2018,Forre2020},  it has been known for some time that the leading-order nondipole correction is given by the effect of the magnetic field component of the laser pulse on the electric dipole-induced motion of the electron along the laser polarization direction. As detailed in Ref.~\cite{JensenPRA2020} this interaction is also dominant in intense midinfrared fields and the associated Hamiltonian is called the nondipole SFA Hamiltonian. Equation \eqref{condition} specifies  the condition on $\omega$ for the applicability of that approach for a given field strength. This approach was recently applied for laser-assisted scattering~\cite{JensenJPB2020} and  photoelectron momentum distributions~\cite{Lund2021}, in the latter case, reaching conclusions regarding nondipole-induced shifts of momentum distributions that are confirmed by time-dependent Schr\"odinger equation simulations~\cite{Lein2021}.

\subsection{Nondipole strong-field-approximation Hamiltonian approach}
\begin{figure}
\includegraphics[width=0.450\textwidth]{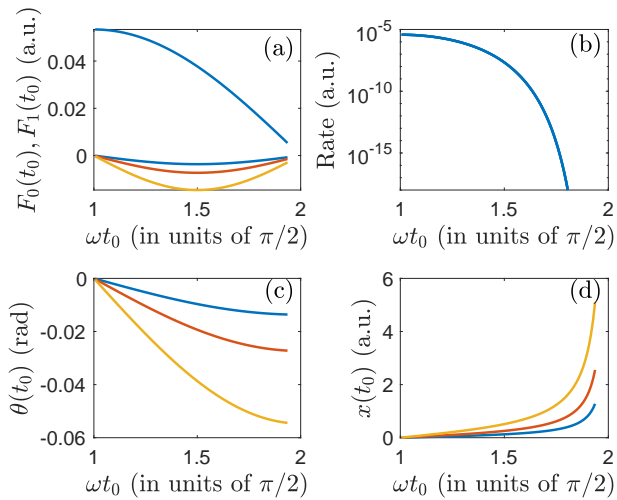}
\caption{(a) Electric field $F_0(t_0)$ [\eq{F0}, positive values] and $F_1(t_0)$ [\eq{F1}] in a half cycle starting from the peak of $F_0(t_0)$ as a function of ionization time $t_0$. The fields $F_1(t_0)$ are for the wavelengths 1600 (blue), 3200, (red) and 6400 (yellow) nm and they have been multiplied by a factor of 10 for clarity.  The minima in $F_1(t_0)$ increase in absolute magnitude for increasing wavelength. The field strength $F_0$  corresponds to an intensity of $10^{14}$ W/cm$^2$. (b) Ionization rate  [\eq{rate}] for $F_0(t_0)$ in (a) and $I_p=0.5$ a.u. (c)  Angle $\theta(t_0)$ [\eq{theta}] between the instantaneous direction of the field $\tilde{\bm F}(t_0)$ [\eq{Fprime}] and the polarization direction, for the fields in (a). The amplitude of $\theta(t_0)$ increases with increasing wavelength. (d) Position $x(t_0)$ [\eq{nondipoleexit}] of the exit point in the {\it propagation} direction for the fields in (a). The  amplitude of $x(t_0)$ increases with increasing wavelength.}
\label{fig1}
\end{figure}

Consider a laser that is linearly polarized along the $z$ direction and propagates along the $x$ direction.  For simplicity the pulse is assumed to contain sufficiently many cycles, say 10 or more,  such that the variation of the envelope can be neglected compared with the variation in the carrier.  The Hamiltonian with full inclusion of retardation reads 
\be
H= (\bm p + \bm A(\eta))^2/2 + V(\bm r),
\ee
with $V(\bm r)$ the effective single-electron potential and $\eta= \omega t - x \omega/c$. The vector potential is expanded to first order in $1/c$, 
\be
\bm A(\eta) = \bm A_0(t) + \bm A_1(x,t),
\ee
with $\bm A_0(t) = \bm A(\eta)|_{\eta = \omega t}$ and $\bm A_1(x,t) = - (\omega x/c) \partial_\eta \bm A(\eta) |_{\eta=\omega t}$. The nondipole SFA Hamiltonian reads~\cite{JensenPRA2020}
\be
H= (\bm p + \bm A_0(t))^2/2 + \bm A_0(t) \cdot \bm A_1(x,t)  + V(\bm r).
\label{H1}
\ee
The nondipole interaction term can be rewritten as 
\be
\label{ND1}
\bm A_0(t) \cdot \bm A_1(x,t) = (\dot {\bm r}_\text{D} \times \bm B_1(t)) \cdot \bm r,
\ee
where $\dot {\bm r}_\text{D} =\bm A_0(t)$ is the velocity of the electron in its electric-dipole-induced motion and where $\bm B_1(t)=\nabla \times \bm A_1(x, t)$ is the leading-order magnetic field of the laser pulse. Hence, the nondipole  term describes the interaction of the electric-dipole-induced motion with the magnetic field. It is noted that the nondipole term on the r.h.s.~of \eq{ND1} is on the form of an electric dipole operator with an effective field given by 
\be
\bm F_1(t) = \dot {\bm r}_\text{D} \times \bm B_1(t),
\label{Fx}
\ee 
and directed along the laser propagation direction as per the cross product on the r.h.s.~of \eq{Fx}. A unitary transformation brings the entire Hamiltonian of \eq{H1} on the length gauge form and the result reads
\be
H = \frac{p^2}{2} + \tilde{\bm F}(t) \cdot \bm r + V(\bm r),
\label{H}
\ee
with a nondipole-modified effective field
\be
\tilde{\bm F}(t) = \bm F_0(t) + \bm F_1(t).
\label{Fprime}
\ee
For a vector potential on the form $ \bm A(\eta) = \hat{\bm z} A_0 \cos(\eta)$, the electric dipole field $\bm F_0(t) = - \partial_t \bm A_0(t)$ reads
\be
\bm F_0(t) = \hat{\bm z} F_0(t)=  \hat{\bm z} F_0 \sin (\omega t),
\label{F0}
\ee
and the nondipole term that follows from the above considerations reads
\be
\bm F_1(t) = \hat{\bm x} F_1(t) = \hat{\bm x} \frac{F_0^2}{\omega c} \sin(\omega t) \cos(\omega t) =  \hat{\bm x} \frac{F_0^2}{2 \omega c} \sin(2\omega t).
\label{F1}
\ee
Equation \eqref{H}  allows the consideration of nondipole terms in adiabatic tunneling ionization, for frequencies fulfilling \eq{condition},  due to the presence of the effective field $\tilde{\bm F}(t)$ given by  \eq{Fprime}. This effective field has the dipole component along the polarization direction ($\hat{\bm z}$) and the nondipole term along the propagation direction ($\hat{\bm x}$). To illustrate typical $\bm F_0(t)$ and $\bm F_1(t)$ fields, it suffices to consider a quarter of a period from the peak of $\bm F_0(t)$ and  Fig.~\ref{fig1}(a) shows an example of their magnitudes  for near- to midinfrared wavelengths.  The oscillation at $2\omega$ in \eq{F1} explains the factor of $1/2$ in \eq{condition}.

\subsection{Strict adiabatic limit}
In this section, the strict adiabatic limit is considered. In this limit  the time of ionization, to be denoted by $t_0$, and therefore the field strength of the laser pulse are considered as parameters, and the tunneling is assumed to occur at an instantaneous value of the field. This means that the process takes place for finite, small $\omega$ and not too high $F_0$ to conform with the restrictions set by \eq{condition}.  It may be helpful to consider some typical numerical values for the quantities that enter \eq{condition}. For example, at an intensity of $10^{14}$ W/cm$^2$  the l.h.s.~of \eq{condition} equals $3.9 \times 10^{-4}$, and for typical atoms the r.h.s.~is around 0.2. The angular frequencies for 3200 nm and 6400 nm light are $1.4 \times 10^{-2}$ and  $7.1 \times 10^{-3}$, respectively, which are both within the $\omega$-range specified by \eq{condition}.

The instantaneous magnitude of the field at the time of ionization $| \tilde{\bm F}(t_0) | = \sqrt{F_0(t_0)^2 + F_1(t_0)^2}  \simeq | \bm F_0(t_0) |$, equals the magnitude of the electric dipole part of the field to order $1/c$. Accordingly, to order $1/c$, the tunneling rate to exponential accuracy is unaffected by the nondipole correction and is given by the conventional static-field result~\cite{LL}
\be
\Gamma (t_0) \propto \exp\left(- \frac{2 \kappa^3}{3 | \tilde{\bm F}(t_0)|}\right) \simeq \exp\left(- \frac{2 \kappa^3}{3 | \bm F_0(t_0)|}\right).
\label{rate}
\ee
Figure \ref{fig1}(b) illustrates this function.  While the ionization rate peaks at the peak of $\bm F_0(t_0)$, it is known that rescattering trajectories, crucial for much of strong-field physics, are born over a relatively large interval. For example in Ref.~\cite{Lin2009}, rescattering trajectories with $1.044 \leqslant \omega t_0/(\pi/2) \leqslant 1.278$ (for the present choice of $\bm F_0(t)$)  were considered. So even though the ionization rate at, say,  $ \omega t_0/(\pi/2) \simeq 1.278$ is  $\sim 1/3$ of the rate at $\bm F_0(t_0)$ extrema, the corresponding trajectories play a role in dynamics, e.g.,  at high final kinetic energies. Since the relative strength of the $\bm F_1(t_0)$ correction increases with time from the field maximum [Fig.~\ref{fig1}(a)], the importance of the nondipole correction  plays a relatively larger role for rescattering trajectories than for direct electrons. A comparison of Figs.~\ref{fig1}(a) and \ref{fig1}(b)  shows that ionization at the peak of $\bm F_1(t_0)$ is suppressed compared to ionization at times close to the maximum of $\bm F_0(t_0)$.

In contrast to $\Gamma(t_0)$,  the tunnel exit points are affected by $\bm F_1(t_0)$ to order $1/c$. The instantaneous direction of the electric field makes an angle, $\theta(t_0)$, with the polarization axis due to the component along the propagation direction, $\tan (\theta(t_0)) = F_1(t_0)/F_0(t_0) = F_0 \cos (\omega t_0)/(\omega c)$, and since $\theta(t_0)$ is small  
\be
\theta(t_0) = \frac{F_0 \cos (\omega t_0)}{\omega c}. 
\label{theta}
\ee
The function $\theta(t_0)$ is illustrated in Fig.~\ref{fig1}(c). In the electric dipole approximation, the tunnel exit point can be accurately obtained by working in parabolic coordinates~\cite{Pfeiffer2012,Nikolay2012}, since the problem involving the sum of the Coulomb potential and the static-field interaction can be variable-separated in these coordinates. Here, to illustrate how the nondipole term shifts the tunnel exit point it suffices to consider the field direction model~\cite{Nikolay2012}.  The component of the exit point in the polarization direction (opposite to the direction of the instantaneous field) at the time of ionization can be estimated by
\be
\label{zexit}
\bm r^\text{D}(t_0)  \simeq  - \hat{\bm z} \frac{I_p}{F_0(t_0)},
\ee 
and it constitutes the exit point in the electric dipole approximation as indicated by the superscript D. In the approach including nondipole effects, the time-dependent exit points are estimated by $\theta(t_0)$-dependent projections and to order $1/c$ the result reads
\be
\label{nondipoleexit}
\bm r(t_0) = -\hat{\bm z} \frac{I_p}{F_0(t_0)} - \hat{\bm x} \frac{I_p}{F_0(t_0)} \theta(t_0).
\ee
Equation \eqref{nondipoleexit} shows that the nondipole term induces a nonvanishing exit point along the $\hat{\bm x}$ propagation direction and Fig.~\ref{fig1}(d) illustrates typical magnitudes. In the time-of-birth window for rescattering trajectories ($1.044 \leqslant \omega t_0/(\pi/2) \leqslant 1.278$~\cite{Lin2009}), $x(t_0)$ may attain values of the order of 10 $\%$ of an atomic unit, and this value increases with wavelength as seen from \eq{theta}. 

Since the instantaneous field direction is not exclusively along the polarization direction, the momenta at the tunnel exit are also affected by the nondipole term.
\begin{figure}
\includegraphics[width=0.450\textwidth]{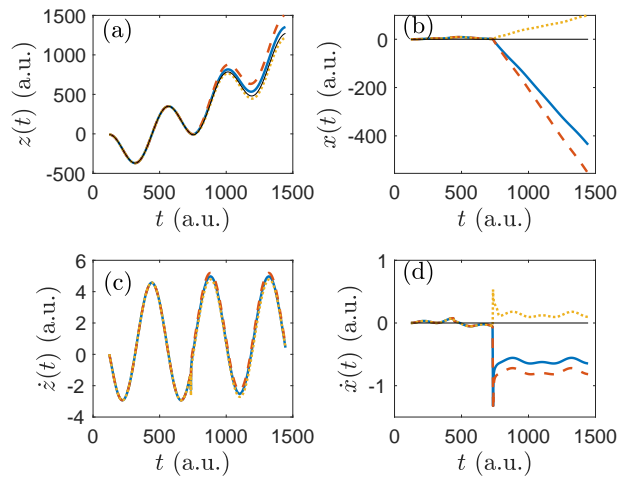}
\caption{Illustration of the motion of an electron in a laser field including nondipole effects and a Coulomb potential with unit attractive charge. Position in the (a) polarization and (b) propagation directions. Velocity  in the (c) polarization and (d)  propagation directions as a function of time. The laser wavelength is 3200 nm and peak intensity is $10^{14}$ W/cm$^2$. The propagation time is 3 cycles from the peak of the electric dipole field. The full (blue) curves show the results for the nondipole exit point [\eq{nondipoleexit}] and a nonzero choice of the initial $k_x =I_p/(3c)$, where the distribution peaks in the propagation direction~\cite{JensenPRA2020}. 
The dashed (red) curves show results for a dipole exit point and initial momentum as for the full, blue curve. The dotted (yellow) curves show results for the electric dipole exit point [\eq{zexit}] and no initial momentum. The thin full curves show result for the electric dipole exit point, no initial momentum and propagation in the electric dipole field. The trajectories are started at $\omega t /(\pi/2) =1.108$ and $I_p=0.5$ a.u..}
\label{fig2}
\end{figure}
If the distribution of the momenta at the tunnel exit is modelled by a Gaussian transverse to the instantaneous field direction $w(k_\perp) \propto \exp(-\kappa k_\perp^2/F_0)$~\cite{Delone1991},  as is typically done in tunneling-based approaches~\cite{Nikolay2016,Ma2021}, the momenta along the propagation and polarization directions at tunneling may be easily modelled. For the asymptotic momentum $k_x$, the momentum at tunneling is $v_x(t_0) = k_x\cos(\theta(t_0)) \simeq k_x$ in the propagation direction. In the polarization direction, one obtains $v_z(t_0)= -k_x \sin(\theta(t_0)) \simeq -k_x \theta(t_0)$ for a given $k_x$ from the distribution $w(k_x)$. In this sense, the nondipole term induces a component along the longitudinal polarization direction, which is absent in the electric dipole approach in the strict adiabatic limit.

The sensitivity of the electron dynamics to the nondipole-induced changes in the tunnel exit point and momentum at the tunnel exit is considered by classical trajectory calculations with full retardation. Figure \ref{fig2} shows an example for an electron initially in the hydrogenic 1s state for a set of laser parameters similar to those used in an experiment reporting nondipole effects in the photoelectron momentum distribution~\cite{Keller2014} (see the caption of Fig.~\ref{fig2} for laser parameters and initial conditions). The excursion in the propagation direction in the nondipole induced figure-of-eight motion of the free electron is $\simeq 1$ and therefore the parameters are at the onset of low-frequency nondipole effects~\cite{Reiss2008,Reiss2014}. Figure \ref{fig2} shows that there are nondipole-induced changes of the time-dependent positions and velocities along both the polarization and propagation directions due to the nondipole-induced modifications at the tunnel exit. In the example considered in the figure, the inclusion of both the nondipole-induced modification of the longitudinal  momentum at the tunnel exit and the modification of the tunnel exit point itself leads to a modification of the dynamics that is different than that obtained by just considering the nondipole-induced spatial displacement of the tunnel exit point. The nondipole-induced changes are most clearly seen in the propagation direction [Figs.~2(b) and 2(d)]. The abrupt changes in the trajectories at a propagation time of around 740 a.u. are due to a revisiting of the core region by the electron, i.e., the electron passes the Coulomb singularity very closely.  While such dynamics depend on the initial conditions of each individual trajectory, the main point remains clear: The final momenta will be affected by the nondipole-induced modifications at the tunnel exit and therefore these effects also need be taken into account in the interpretation of data, such as, e.g., holographic and laser-induced diffraction patterns.

\subsection{Tunnel exit point and momentum at time of ionization from nondipole SFA considerations}
The results of this section describe nonadiabatic corrections as captured by the nondipole Volkov phase. It is shown that the findings regarding nondipole-induced modifications of initial momentum at tunneling and exit points also follow from a consideration using a nondipole SFA approach in the adiabatic limit. The results follow from a nondipole extension of the dipole results discussed, e.g., in Refs.~\cite{Popruzhenko2014,Liu2016,Pisanty2016}.  

The starting point in this analysis is a consideration of the action phase associated with strong-laser-field ionization in the nondipole-SFA-Hamiltonian approach~\cite{JensenPRA2020,Lund2021}
\be
\label{S}
S(t)=- \int_{- \infty}^t (\bm k + \tilde{\bm A}(t'))^2/2 \,dt' + I_p t.
\ee 
Here the factor $I_p t$ comes from the time evolution of the initial bound state and the momentum $\bm k$ is the asymptotic momentum of the outgoing electron, i.e., the momentum that could be measured at a detector.  The integral term comes from the nondipole Volkov phase with the nondipole-modified vector potential
\be
\tilde{\bm A}(t) = \bm A_0(t) + \bm A_{M}(t).
\ee
Here $\bm A_0(t) = \hat {\bm z} A_0 \cos(\omega t)$, as before, and $\bm A_{M}(t) = \hat{\bm x} \frac{A_0^2}{2 c} \cos^2(\omega t)$ is a $1/c$ nondipole correction to the vector potential, which accounts for the dominant effect of the magnetic field component of the electromagnetic field along the dipole induced electron trajectories. Clearly, $\bm F_1(t) = -\partial_t  \bm A_{M}(t)$ with $\bm F_1(t)$ the nondipole correction field of \eq{F1}. Note that the atomic potential does not enter the action of \eq{S} explicitly, the potential is only reflected through the presence of $I_p$. This impedes a dependence of the exit point on the spatial variation of the atomic potential. Therefore, the nondipole SFA action phase approach can not describe situations where the exit to the continuum occurs at distances where the atomic potential is not negligible compared to the dipole interaction of the laser field~\cite{Pfeiffer2012}. In the strict adiabatic limit, such potential effect can be accounted for by improving the approximation leading to \eq{zexit} as described in Refs.~\cite{Pfeiffer2012,Nikolay2012}. Effects of the ionic potential on the outgoing electron including those imparted on the initial conditions can be taken into account in  SFA-related  techniques \cite{CarlaFaria2020}.

In the nondipole SFA approach, the notions of (initial) momentum at tunnel exit and spatial tunnel exit point come from a consideration of stationary phase of the action of \eq{S} w.r.t. variation in time, i.e., they come from interpretation of the consequences of the condition, $\partial_t S(t) = 0$. The latter equation has complex times $t_s = t_0 + i \tau_0$ as solutions. These times solve
\be
\label{dsdt}
(\bm k + \tilde{\bm A}(t_s))^2 + \kappa^2= 0.
\ee
Inserting the expression for $\tilde{\bm A}(t)$ in \eq{dsdt} leads to the saddle point solutions
\be
\label{ts}
\cos(\omega t_s) =\frac{\omega}{F_0}   \Big\{ \frac{-k_z \pm i [k_\perp^2 + \kappa^2 + (k^2 + \kappa^2)k_x/c]^{1/2}}{1+k_x/c} \Big\},
\ee
with $k_\perp^2 = k_x^2+k_y^2$ the momentum transverse to the polarization direction. The real time $t_0$ is the time of ionization as in the previous Sec.~II.B. In the dipole approximation, the factor $k_x/c$ goes to zero. 

With the complex $t_s$ at hand, the tunnel exit point is found from $\text{Re} [\bm r(t_0)] = \text{Re} [ \int_{t_s}^{t_0} (\bm k + \tilde{\bm A} (t)) d t] $. The interpretation of this integral is that the electron moves through the effective tunneling barrier  in imaginary time from $t_s$ to $t_0$, i.e., through the complex time interval $t_0 -t_s= -i \tau_0$. An evaluation of the integral gives the tunnel exit point at $t_0$,
\ba
\label{r0}
\text{Re}[\bm r(t_0)] &=& \hat {\bm z} \frac{F_0 (t_0)}{\omega^2} [1 - \cosh(\omega \tau_0)]  \\ \nonumber
&+& \hat {\bm x} \frac{F_0(t_0)}{4\omega^3 c} F_0  \cos(\omega t_0) [1- \cosh(2 \omega \tau_0))].
\ea
Here, the nondipole modification of the exit point is the component after the laser propagation direction, $\hat{\bm x}$.

To obtain the momentum at the tunnel exit, consider the condition $\text{Im} [\partial_t S(t)] =0$. A calculation then shows that the asymptotic longitudinal momentum at the detector, $k_z$, can be expressed as
\be
\label{kz}
k_z = - A_0(1 + \frac{k_x}{c}) \cos(\omega t_0) \cosh(\omega \tau_0).
\ee
This relation means that, at the instant of tunneling, the kinematic momentum $\bm v(t_0) = \bm k + \tilde{\bm A}(t_0)$, reads
\ba
\label{v_init}
\bm v(t_0) &=& \hat{\bm z} \frac{F_0}{\omega} \cos(\omega t_0) [1 - (1 + \frac{k_x}{c}) \cosh(\omega \tau_0) ]
\\ \nonumber
&+& \hat{\bm x} [k_x + \frac{F_0^2}{2\omega^2 c} \cos^2(\omega t_0)] + \hat{\bm y} k_y.
\ea
The component along the propagation direction ($\hat{\bm x}$) carries, for fixed intensity, a $\omega^{-2}$ scaling. Such a scaling was considered in Ref.~\cite{Willenberg2019} for gaining agreement between experimental data obtained in the midinfrared regime and results of a two-step semiclassical simulation including nondipole terms in the propagation of the classical equations of motion after tunneling.

While the expressions of Eqs.~\eqref{r0} and \eqref{v_init} are generally valid within  the nondipole SFA approach, a connection to the results obtained in the strict adiabatic limit requires a consideration of the limit of small $\omega$ in combination with $\omega$'s fulfilling \eq{condition}; see the beginning of Sec.~II.B for typical numerical values for the quantities entering \eq{condition}.
To proceed, first the relation between $t_0$ and $\tau_0$ is considered. This relation is obtained by using  that $\text{Re} [\partial_t S(t)] = 0$, i.e., the real part of \eq{dsdt} should be zero. This requirement gives, in combination with the use of \eq{kz},
\ba
\label{t0tau0}
k_\perp^2 + \kappa^2 &=& \frac{F_0(t_0)^2}{\omega^2} (1 + \frac{k_x}{c}) \sinh^2(\omega \tau_0) \\ \nonumber
&-& \frac{F_0^2}{\omega^2} \frac{k_x}{c} \cos^2(\omega t_0) \cosh^2(\omega \tau_0).
\ea
The relation between $t_0$ and $\tau_0$ then reads
\be
\sinh^2(\omega \tau_0) = \frac{(\omega^2/F_0^2)(k_\perp^2 + \kappa^2)+ (k_x/c) \cos^2(\omega t_0)}{(1+k_x/c) \sin^2(\omega t_0) - (k_x/c) \cos^2(\omega t_0)},
\ee
where the $k_x$-dependent factors on the r.h.s. account for the nondipole modification. In the limit of small $\omega$, and to first order in $1/c$ it then follows that the dominant contribution to the imaginary part of $t_s$ is given by
\be
\label{lim1}
\cosh(\omega \tau_0) = 1 + \frac{\omega^2(k_\perp^2 + \kappa^2)}{2 F_0(t_0)^2 } (1-k_x/c),
\ee
and therefore
\be
\label{lim2}
\cosh(2\omega \tau_0) = 1 + \frac{2\omega^2(k_\perp^2 + \kappa^2)}{F_0(t_0)^2 } (1-k_x/c).
\ee
\begin{figure}
\includegraphics[width=0.450\textwidth]{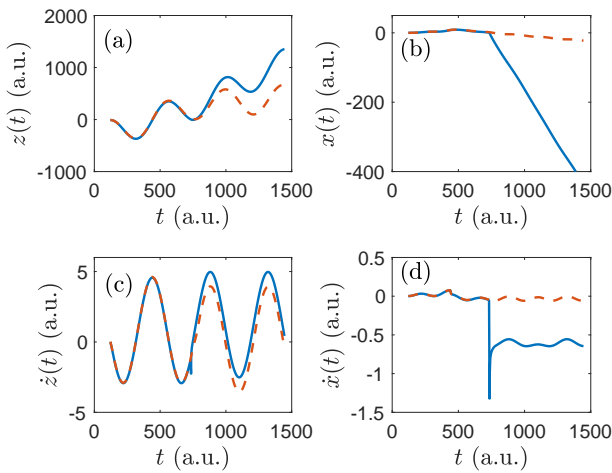}
\caption{Illustration of the motion of an electron in a laser field including nondipole and nonadiabatic effects, and a Coulomb potential with unit attractive charge. Position in the (a) polarization and (b) propagation directions. Velocity  in the (c) polarization and (d)  propagation directions as a function of time. The laser parameters and propagation times is as in Fig.~2.  The full (blue) curves are the same as in Fig.~2 and show the results for the nondipole exit point [\eq{nondipoleexit}] and a nonzero choice of the initial $k_x =I_p/(3c)$, where the distribution peaks in the propagation direction~\cite{JensenPRA2020}. The dashed (red) curves show results when the nonadiabatic corrections in the initial position [\eq{r02}] and initial velocity [\eq{v_init2}] are included for $k_y=0$. The trajectories are started at $\omega t /(\pi/2) =1.108$ and $I_p=0.5$ a.u..}
\label{fig3}
\end{figure}
So, in this limit, and to first order in $1/c$, \eq{r0} for the tunnel exit point reduces to 
\ba
\label{r02}
\text{Re}[\bm r(t_0)] &=& -\hat {\bm z} \frac{E_\perp + I_p}{ F_0 (t_0)} (1-k_x/c)   \\ \nonumber
&-& \hat {\bm x} \frac{E_\perp + I_p}{F_0(t_0)} \theta(t_0). 
\ea
Here $E_\perp = k_\perp^2/2$, and $\theta(t_0)$ is defined in \eq{theta}. It is readily seen that the expression in \eq{r02} reduces to the one contained in \eq{nondipoleexit} in the limit of small longitudinal and transverse momenta. Likewise, it is found from \eq{v_init} and \eq{lim1} that for small $\omega$, the velocity at the exit point at the time of tunneling reads
\ba
\label{v_init2}
\bm v(t_0) &=& - \hat{\bm z}  \theta(t_0) \left( k_x + \frac{c \omega^2(E_\perp + I_p)}{ F_0(t_0)^2} \right)\\ \nonumber
&+& \hat{\bm x} [k_x + \frac{F_0^2}{2\omega^2 c} \cos^2(\omega t_0)] + \hat{\bm y} k_y.
\ea
 This expression for the momentum along the polarization ($\hat{\bm z}$) direction at the tunnel exit  reduces to the one of the strict adiabatic limit of Sec.~II.B when $(\omega/F_0)^2$ becomes small.

From the above results it is seen that in both the strict adiabatic limit and in the case of the nonadiabatic approach with the nondipole Volkov phase, the nondipole correction increases with decreasing $\omega$ as $1/\omega$ due to the form of $\theta(t_0)$ [\eq{theta}].  This behavior as function of $\omega$ is different than the dipole result. In the dipole case, the $k_x$-term in the $\hat{\bm z}$ direction in \eq{v_init2} is absent and the  longitudinal momentum at the tunnel exit is proportional to $\omega$ as $\omega\rightarrow 0$ as seen from the last factor along  the $\hat{\bm z}$ direction in \eq{v_init2}. In considering these limits, the condition imposed by \eq{condition} should of course be kept in mind. The requirement of a small quiver velocity, $F_0/\omega$, compared to the speed of light, $c$, means that $\omega$ can not be chosen arbitrarily small. 

Figure \ref{fig3} shows an example illustrating, in addition to the nondipole effects of the strict adiabatic limit, the effect of the nonadiabatic terms in the initial conditions, i.e., the effect of including nonadiabatic shifts as in \eq{v_init2} and \eq{r02}. In the figure, the full (blue) curves are as in Fig.~\ref{fig2}. That is, these results are for the nondipole initial conditions in the strict adiabatic limit of Sec.~II.B, and propagation in the full nondipole field and Coulomb potential. The dashed (red) curves include the nonadiabatic terms in the initial conditions. The main reason for the difference in the results in the particular realization of Fig.~\ref{fig3} is due to the extra term in the initial velocity in the polarization ($\hat{\bm z}$) direction in \eq{v_init2} as compared to the initial nondipole-modified velocity in the strict adiabatic limit. This extra nonadiabatic term means that the electron does not traverse the Coulomb singularity at a propagation time of around 740 a.u. as it does without this additional nonadiabatic initial off-set. A simulation performed under the same conditions but for 6400 nm light showed no effect of including the nonadiabatic terms in the initial conditions. This 6400 nm case is still within the frequency range defined by \eq{condition} and the smaller $\omega$ leads to the decrease in the nonadiabatic off-sets of the initial conditions.

\section{Conclusion and outlook}

In summary, a nondipole correction of order $1/c$ associated with the dipole-induced motion in the magnetic field component  of the laser pulse was considered. The interaction leads to a term that can be interpreted as an additional electric field component directed along the propagation direction of the laser pulse. In this case, the ionization rate  is unaffected by the nondipole contributions to exponential accuracy. The tunnel exit and momentum at the tunnel exit, however, are affected by the nondipole correction. For example, nonvanishing initial momenta, which increase with decreasing $\omega$, are predicted along the laser polarization direction.  Such initial momenta are vanishing in tunneling theory when nondipole effects are neglected. It was illustrated that the nondipole-induced changes in the initial conditions lead to changes in the electron dynamics and the final momentum distributions.  The nonadiabatic modifications of the initial conditions following a treatment with a  nondipole SFA approach were also considered. The results of this work outline how to incorporate nondipole effects in the initial tunneling step and associated phasespace initial conditions and hence open the exploration of nondipole strong-field and attosecond physics based on familiar semiclassical two-step ionization and three-step recombination models.

\begin{acknowledgments}
Useful discussions with Andrew S. Maxwell, Henrik Stapelfeldt, Simon V. B. Jensen, Mads M. Lund, and Thomas Hansen are acknowledged. This work was supported by the Independent Research Fund Denmark (Grant No. 9040-00001B and 1026-00040B).
\end{acknowledgments}

%



\end{document}